\documentclass[10pt]{revtex4}
\usepackage{amssymb}
\usepackage{latexsym}
\usepackage{epsfig}
\begin{document}                             

\title{Sounds of instability from generalized QCD ghost dark energy}

\author{Esmaeil Ebrahimi$^{1,2}$ \footnote{eebrahimi@uk.ac.ir} and Ahmad  Sheykhi$^{2,3}$ \footnote{
asheykhi@shirazu.ac.ir} }
\address{$^1$ Department of Physics, Shahid Bahonar University, PO Box 76175, Kerman, Iran\\
          $^2$ Research Institute for Astronomy and Astrophysics of Maragha (RIAAM), Maragha, Iran\\
          $^3$ Physics Department and Biruni Observatory, College of
Sciences, Shiraz University, Shiraz 71454,
          Iran}

\begin{abstract}
We investigate about the stability of generalized QCD ghost dark
energy model against perturbations in the FRW background. For this
purpose, we use the squared sound speed $v_s^2$ whose sign
determines the stability of the model. We explore the stability of
this model in the presence/absence of interaction between dark
energy and dark matter in both flat and non-flat geometry. In all
cases we find almost a same result. Based on the square sound
speed analysis, due to the existence of a free parameter in this
model, the model is theoretically capable to lead a dark energy
dominated stable universe. However, observational constraints rule
out such a chance. In conclusion, we find evidences that the
generalized ghost dark energy might can not lead to a stable
universe favored by observations at the present time.
\end{abstract}

 \maketitle
\section{Introduction}
Seeking a convincing explanation for the present acceleration of
the cosmic expansion \cite{Rie}, a variety of models have been
proposed in the literature. These models should be consistent with
observational evidences supporting the acceleration phase of the
universe expansion \cite{cmb1,cmb2,sdss1,sdss2}. All of the models
can be categorized in two different groups the dark energy (DE)
\cite{wetter,ratra,chiba,armend1,armend2,cald,tachyon,HDE,ADE,nederev}
and the modified gravity \cite{capoz,carroll,DGP,sheywang} models.
Both of these approaches are extensively investigated in the
literature during the past decade. In this paper we are interested
in DE approach. DE models are based on the assumption that the
correct theory of gravitation in large scale is the Einstein's
general relativity (GR). In this approach it is assumed that there
exist an exotic type of energy which its equation of state
parameter is negative and push the universe to accelerate.

One main problem in solving the DE problem is introducing new
degrees of freedom which may lead to inconsistencies. Thus
avoiding such problem we are more interested in DE models, based
on already presented degrees of freedom in physics. One
interesting model of this category is the so called ghost dark
energy (GDE) model proposed recently \cite{Urban,Ohta,CaiGhost}.
Seeking a solution to $U(1)$ problem, the so-called Veneziano
ghost has been proposed in the low energy effective QCD where they
are completely decoupled from the physical sector
\cite{kawar,witten,rosen,nath}. However, they contribute to the
vacuum energy in curved space or time-dependent background. In
\cite{Ohta} the authors discussed that this vacuum energy can play
the role of DE. This contribution to the vacuum energy in curved
space-time is proportional to $\Lambda^3_{QCD} H$,
 where $H$ is the Hubble parameter and $\Lambda^3_{QCD}$ is QCD mass scale.
 With $\Lambda_{\rm QCD}\sim 100 MeV$ and $H \sim 10^{-33}eV$ ,
$\Lambda^3_{\rm QCD}H$ gives the right order of magnitude $\sim
(3\times10^{-3}eV)^4$ for the observed DE density \cite{Ohta}.
Following this interesting coincidence, various aspects of GDE
have been studied
\cite{shemov,ebrins,shemoveb,shebagh,ebrbd,rozas,karami,khodam1,khodam2,feng}.

In \cite{zhi}, the authors discussed that the contribution of the
Veneziano QCD ghost field to the vacuum energy is not exactly of
order $H$ and a subleading term $H^2$ appears due to the fact that
the vacuum expectation value of the energy-momentum tensor is
conserved in isolation \cite{mig}. They argued that the vacuum
energy of the ghost field can be written as $H+O(H^2)$, where the
subleading term $H^2$ in the GDE model might play a crucial role
in the early evolution of the universe, acting as the early DE.
Based on this idea people considered the role of this version of
the Veneziano ghost field energy to the DE problem and tried to
see if there exist a better agreement between this latter model
and the observations. We call this model as generalized ghost dark
energy (GGDE). In this model the energy density is written in the
form $\rho_D= \alpha H+\beta H^2$, where $\beta$ is a constant
\cite{caighost2}. It was shown \cite{caighost2} that taking the
subleading term $H^2$ into account can give better agreement with
observational data compared to the usual GDE. The GGDE has
attracted a lot interest recently \cite{ebrsheyggde,karami}. It
was shown \cite{ebrsheyggde,caighost2}, that the subleading term
$H^2$ in the energy density has negative contribution compared to
the leading term $H$. This result first pointed out in
\cite{caighost2} for flat FRW universe and then confirmed for
non-flat universe \cite{ebrsheyggde}.

Based on all mentioned above, in this paper we would like to
investigate the effects of the subleading term on the background
perturbations and stability of the GGDE model. It is worth
mentioning that in \cite{ebrins}, we found that the GDE is
instable against perturbations and a stable DE dominated universe
cannot be achieved in such a model.

From observations we know that our universe is in a DE dominated
phase. Thus any viable DE model should result a stable DE
dominated universe. One simple way to check such a stability for
any new DE model is to discuss the behavior of the square sound
speed ($v_s^2 = dp/d\rho$) \cite{peebleratra} in a DE dominated
universe. The sign of $v_s^2$ plays a crucial role in determining
the stability of the background evolution. If $v_s^2<0$, it means
that we have the classical instability of a given perturbation. In
contrast $v_s^2>0$, leaves chance for greeting a stable universe
against perturbations. However, this does not enough insight to
say the model is surely stable but at least can show sounds of
instability in the model. This approach has been used for
exploring some DE models. For example in \cite{myung1,myung2} the
authors investigated the behavior of the square sound speed for
holographic DE as well as the agegraphic DE models and found both
of these models are instable against background perturbations.
Also it was shown that chaplygin gas and tachyon DE have positive
squared speeds of sound with, $v_s^2= -w$, and thus they are
supposed to be stable against small perturbations
\cite{gorini,sandvik}. The stability of the original GDE was
studied in \cite{ebrins}, and we found the GDE is not capable to
result a stable DE dominated universe. The issue was investigated
for flat, non-flat in the presence/absence of the interaction
between DE and dark matter. Also, a same procedure was considered
in \cite{saaidi} to show the stability of the GDE in the chameleon
Brans-Dicke theory.

In this paper our aim is to discuss the chance of the GGDE model
in resulting to DE dominated universe. In recent years several
signals have been detected, implying a small interaction between
DE and DM is possible. As an instance, observational evidences
provided by the galaxy cluster Abell A586 supports the interaction
between DE and DM \cite{interact1}. Also evidences from CMB and
also supernova measurements of the cubic correction to the
luminosity distance favor a positively curved universe
\cite{nonflat2,nonflat3}. Due to the above observational evidences
we consider the stability of GGDE model in the presence of
interaction in both flat and non-flat background.

This paper is organized as follows. In the next section, we review
the GGDE model in flat universe and discuss its instability
against perturbation. In section \ref{II}, we explore the
stability of the GGDE model in the presence of the interaction
term in flat universe. Section \ref{III} is devoted to stability
of the GGDE model in a interacting, non-flat universe. We
summarize our results in section \ref{sum}.
\section{Stability of non-interacting GGDE model}\label{I}
\subsection{Review of the non-interacting GGDE in flat universe}
At first we briefly discuss the non-interacting GGDE in a flat FRW
background. For flat universe, the first Friedmann equation read
\begin{eqnarray}\label{Fried}
H^2=\frac{8\pi G}{3} \left( \rho_m+\rho_D \right),
\end{eqnarray}
where $\rho_m$ and $\rho_D$ are, respectively, the energy
densities of pressureless matter and DE. The dimensionless energy
density parameters are defined as
\begin{equation}\label{Omega}
\Omega_m= \frac{8\pi G \rho_m}{3 H^2}, \  \   \
 \Omega_D=\frac{8\pi G \rho_D}{3 H^2},
\end{equation}
According to these definitions, the first Friedmann equation
(\ref{Fried}) can be rewritten as
\begin{equation}\label{fridomega}
\Omega_m+\Omega_D=1.
\end{equation}
The conservation equations also read
\begin{eqnarray}
\dot\rho_m+3H\rho_m&=&0,\label{consm}\\
\dot\rho_D+3H\rho_D(1+w_D)&=&0\label{consd},
\end{eqnarray}
which imply that matter and DE components are separately
conserved.

The energy density of GGDE is defined as \cite{caighost2}
\begin{equation}\label{GGDE}
\rho_D=\alpha H+\beta H^2,
\end{equation}
where $\alpha$ is a constant of order $\Lambda_{\rm QCD}^3$ and
$\Lambda_{\rm QCD}$ is QCD mass scale. In the original GDE with
$\Lambda_{\rm QCD}\sim 100MeV$ and $H\sim 10^{-33}eV$ ,
$\Lambda_{\rm QCD}^3 H$ gives the right order of magnitude $\sim
(3\times 10^{-3}\rm {eV})^4$ for the observed DE density
\cite{Ohta}. In the GGDE, $\beta$ is a free parameter which can be
adjusted for better agreement with observations.

Taking the time derivative of relation (\ref{GGDE}), we obtain
\begin{equation}\label{dotrho}
\dot{\rho}_D=\dot{H}(\alpha+2\beta H).
\end{equation}
Also differentiating (\ref{Fried}) with respect to time lead
\begin{equation}\label{hdot}
    \dot{H}=-4\pi G\rho_D(1+u+w_D),
\end{equation}
where $u=\rho_m/\rho_D$. Replacing this relation in (\ref{dotrho})
and also using the continuity equation (\ref{consd}) we get
\begin{equation}\label{prewD}
(1+w_D)[3H-4\pi G(\alpha+2\beta H)]=4\pi G (\alpha+2\beta H).
\end{equation}
Solving the above equation for $w_D$  and noting that
$u=\frac{\Omega_m}{\Omega_D}$ as well as
\begin{equation}\label{rhomaking}
    \frac{4\pi G}{3 H}(\alpha+2\beta
    H)=\frac{\Omega_D}{2}+\frac{4\pi G \beta}{3},
\end{equation}
we obtain
\begin{equation}\label{wD}
w_D=\frac{\xi-\Omega_D}{\Omega_D(2-\Omega_D-\xi)},
\end{equation}
where $\xi=\frac{8\pi G \beta}{3}$. It is clear that this relation
reduce to its respective one in the GDE when $\xi=0$
\cite{shemov}.

It is easy to see that at late times where $\Omega_D\rightarrow
1$, $w_D\rightarrow -1$, which implies that the GGDE in a flat
universe mimics a cosmological constant behavior. Also we find
that $w_D$ of the GDE model cannot cross the phantom divide and
the universe has a de Sitter phase at late time.

Next we turn to the deceleration parameter which is defined as
\begin{equation}\label{q1}
q=-\frac{a\ddot{a}}{\dot{a}^2}=-1-\frac{\dot{H}}{H^2},
\end{equation}
where $a$ is the scale factor. Using Eq. (\ref{hdot}) and the
definition of $\Omega_D$ in (\ref{Omega}) we get
\begin{equation}\label{dotH}
\frac{\dot{H}}{H^2}=-\frac{3}{2}\Omega_D \left(1+u+w_D\right).
\end{equation}
Replacing this relation into  (\ref{q1}), and using (\ref{wD}) we
find
\begin{equation}\label{qfnonint}
q=\frac{1}{2}-\frac{3}{2}\frac{\xi-\Omega_D}{(\Omega_D+\xi-2)}.
\end{equation}
One can easily check that the deceleration parameter in GDE is
retrieved for $\xi=0$ \cite{shemov}. At late time where the DE
dominates ($\Omega_D\rightarrow 1$), independent of the value of
the $\xi$, we have $q=-1$. Besides, taking $\Omega_{D0}=0.72$  and
adjusting $\xi=0.01$, obtain $q_0 \approx-0.34$ for the present
value of the deceleration parameter which is in agreement with
recent observational data \cite{daly}.

The evolution equation of the GGDE can be obtained as
\cite{ebrsheyggde}
\begin{equation}\label{Omegaprime2}
\frac{d\Omega_D}{d\ln a}=-3
\frac{(1-\Omega_D)(\xi-\Omega_D)}{2-\Omega_D-\xi}.
\end{equation}
\subsection{Stability of the model according to the square sound speed parameter}
The parameter we use through this paper as a factor of stability
of the model is the square sound speed $v_s^2$. In classical
theory of perturbation we assume a small fluctuation in the
background energy density and we would like to see if the
perturbation grows or will collapse. In the linear perturbation
regime, the perturbed energy density of the background can be
written as
\begin{equation}\label{pert1}
    \rho(t,x)=\rho(t)+\delta\rho(t,x),
\end{equation}
where $\rho(t)$ is unperturbed background energy density. The
energy conservation equation ($\nabla_{\mu}T^{\mu\nu}=0$) yields
\cite{peebleratra}
\begin{equation}\label{pert2}
\delta\ddot{\rho}=v_s^2\nabla^2\delta\rho(t,x),
\end{equation}
where $v_s^2=\frac{dP}{d\rho}$ is the squared of the sound speed.
Solutions of equation (\ref{pert2}) include two cases of interest.
First when $v_s^2$ is positive  Eq. (\ref{pert2}) becomes an
ordinary wave equation whose solutions would be oscillatory waves
of the form $\delta \rho=\delta\rho_0e^{-i\omega
t+i\vec{k}.\vec{x}}$ which indicates a propagation mode for the
density perturbations. The second case of interest occurs when
$v_s^2$ is negative. In this case the frequency of the
oscillations becomes pure imaginary and the density perturbations
will grow with time as $\delta \rho=\delta\rho_0e^{\omega
t+i\vec{k}.\vec{x}}$. Thus the growing perturbation with time
indicates a possible emergency of instabilities in the background.

Here we would like to obtain the sound speed in a flat FRW
background filled with pressureless matter and GGDE while matter
and DE components are separately conserved. The definition of the
sound speed reads \cite{CaiGhost}
\begin{equation}\label{speeddef2}
    v_s^2=\frac{dP}{d\rho}=\frac{\dot{P}}{\dot{\rho}}=\frac{\rho}{\dot{\rho}}\dot{w}+w,
\end{equation}
where in the last step we have used $P=w\rho$. Using the
conservation equation (\ref{consd}), we obtain
\begin{equation}\label{rhodot}
 \frac{\rho}{\dot{\rho}}=\frac{-1}{3H(1+w_D)}.
\end{equation}
\begin{figure}\epsfysize=5cm
{ \epsfbox{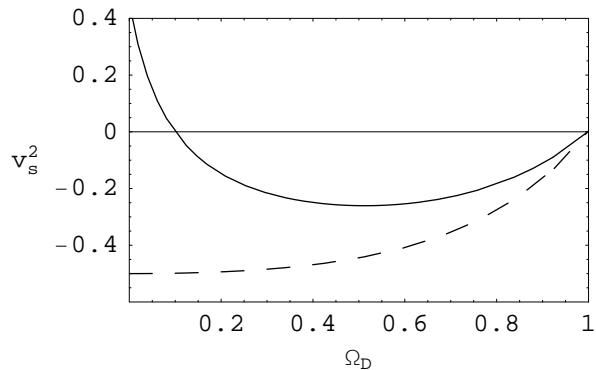}}\caption{Evolution of squared sound speed
$v_s^2$ against $\Omega_D$ for noninteracting GDE (dashed curve)
and GGDE (continues curve) in flat background for $\xi=0.1$.}
\label{fvnin}
\end{figure}
Also taking the time derivative of Eq. (\ref{wD}) yields
\begin{equation}\label{wdot}
    \dot{w}_D=\frac{dw}{d\Omega_D}\dot{\Omega}_D=-\frac{\Omega_D(2\xi-\Omega_D)+\xi(2-\xi)}{\Omega^2_D(2-\Omega_D+\xi)^2}\dot{\Omega}_D.
\end{equation}
Replacing Eqs.(\ref{rhodot}) and (\ref{wdot}) into
(\ref{speeddef2}) and also using (\ref{Omegaprime2}) we get
\begin{equation}\label{v2ni}
    v_s^2=-2\frac{(\xi-\Omega_D)(\Omega_D-1)}{(\Omega_D+\xi)(2-\Omega_D+\xi)^2},
\end{equation}
where we have used $\frac{d}{dt}=H\frac{d}{d\ln a}$. One can
easily see that setting $\xi=0$, this result reduces to its
respective relation in \cite{ebrins}. Having the $v_s^2$ at hand
we are ready to discuss about the stability of perturbations. One
can easily see from (\ref{v2ni}) that $v_s^2$ is negative provided
$\xi>\Omega_D$. One should note here that $v_s^2$ can be positive
up to present time if $\xi$ is limited to suitable values and thus
$\xi$ plays a crucial roles in the stability of perturbations
background. However, according to \cite{caighost2}, the presented
range of parameter $\beta$ does not let us to set $\xi$ to
suitable values. This result indicates that due to the negativity
of the squared sound speed every small perturbation can grow with
time which leads to an instability in the universe. Thus we can
not expect a noninteracting GGDE dominated universe in the future
as the fate of the universe, however to speak about the stability
issue one have to consider other features in the standard theories
of perturbation in cosmology but the negativity of the square
sound speed can be taken as a sign of instability against small
perturbations. The evolution of $v_s^2$ versus $\Omega_D$ is shown
in Fig. \ref{fvnin}.
\section{Stability of interacting GGDE in flat universe}\label{II}
Based on a traditional and historical manner usually people
consider dark matter and DE separately. However, these days
observations detect signals of interaction between DM and DE
\cite{interact1}. Also from theoretical point of view we know that
any conservation law should reflect a symmetry in the Lagrangian
and so far we do not know such a symmetry for DE. Thus, it is
natural to assume the interaction between the two dark components
of the universe.

In order to study the interacting model we consider the energy
balance equations as
\begin{eqnarray}
\dot\rho_m+3H\rho_m&=&Q,\label{consmi}\\
\dot\rho_D+3H\rho_D(1+w_D)&=&-Q\label{consdi},
\end{eqnarray}
where $Q$ represents the interaction term which allows the energy
exchange between the two dark components of the universe. The form
of $Q$ is a matter of choice and we take it as
\begin{equation}\label{Q}
Q =3b^2 H(\rho_m+\rho_D)=3b^2 H\rho_D(1+u),
\end{equation}
with $b^2$  being a coupling constant. Taking the above equations
into account and following a same steps as the previous section we
can find
\begin{equation}\label{wDi}
w_D=-\frac{1}{2-\Omega_D-\xi}\left(1+\frac{2b^2}{\Omega_D}-\frac{\xi}{\Omega_D}\right).
\end{equation}
When $b=0$, $w_D$ reduces to its respective relation in the
absence of interaction. The first interesting point about the
equation of state parameter of the GGDE is that in the interacting
case independent of the interaction parameter, $b$, for $0<\xi<1$,
$w_D$ can cross the phantom line in the future while
$\Omega_D\rightarrow1$.

The deceleration parameter in the presence of an interaction term
can be obtained by substituting (\ref{wDi}) in (\ref{dotH}) and
using (\ref{q1}). We find \cite{ebrsheyggde}
\begin{equation}\label{q3}
q=\frac{1}{2}+\frac{3}{2}\frac{\Omega_D}{(2-\Omega_D-\xi)}
\left(1+\frac{2b^2}{\Omega_D}-\frac{\xi}{\Omega_D}\right).
\end{equation}
Once again it is clear that setting $b=0$, respective relation in
the previous section is retrieved. Finally, we would like to
obtain the evolution equation of DE in the presence of
interaction. It is a matter of calculation to show that
\cite{ebrsheyggde}
\begin{equation}\label{Omegaprime3}
\frac{d\Omega_D}{d\ln a}=3 \Omega_D\left[\frac{1-\Omega_D}{2-
\Omega_D-\xi}\left(1+\frac{2b^2}{\Omega_D}-\frac{\xi}{\Omega_D}\right)-\frac{b^2}{\Omega_D}\right].
\end{equation}
Now we consider the stability of this model by study the sign of
squared sound speed $v_s^2$. From Eq. (\ref{wDi}) we have
\begin{equation}\label{wdoti}
    \dot{w}_D=\frac{dw}{d\Omega_D}{\dot{\Omega}_D}=-\frac{\Omega_D^2+(2b^2-\xi)(2\Omega_D-2+\xi)}
    {\Omega^2_D(2-\Omega_D+\xi)^2}\dot{\Omega}_D
    .
\end{equation}
Also from Eqs. (\ref{consdi}) and (\ref{Q}) one finds
\begin{equation}\label{rrhodot2}
    \frac{\rho}{\dot{\rho}}=-\frac{1}{3H(1+w_D+\frac{b^2}{\Omega_D})}.
\end{equation}
Taking into account above relations, $\frac{d}{dt}=H\frac{d}{d\ln
a}$, and  Eq. (\ref{Omegaprime3}) and replacing these relations
into (\ref{speeddef2}) it is a matter of calculation to show that
\begin{equation}\label{v2in}
    v_s^2=-2\frac{(\Omega_D-1)(\xi-\Omega_D)}{(\Omega_D-2+\xi)^2(\Omega_D+\xi)}
    +2b^2\frac{3\Omega_D-4+\xi}{(\Omega_D-2+\xi)^2(\Omega_D+\xi)},
\end{equation}
\begin{figure}\epsfysize=5cm
{ \epsfbox{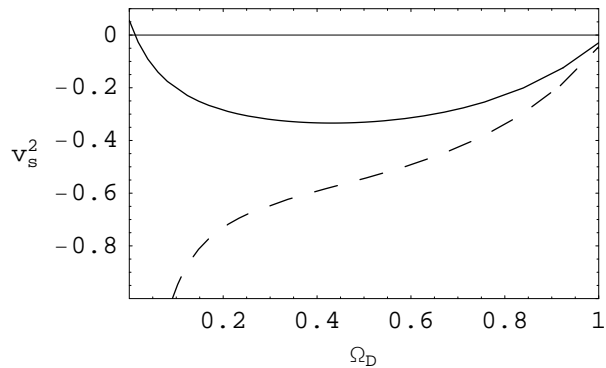}}\caption{This figure shows evolution
of squared sound speed $v_s^2$ versus $\Omega_D$ for interacting
GGDE model. The solid curve corresponds to the GGDE for $b=0.15$
and $\xi=0.1$. Dashed curve is plotted for the original GDE.}
\label{fvin}
\end{figure}
Once again setting $\xi=0$ leads the results of the original GDE
in \cite{ebrins}. The evolution of $v_s^2$ against $\Omega_D$ has
plotted in Fig. (\ref{fvin}) for $b=0.15$ and $\xi=0.1$. This
figure reveals that $v_s^2$ is always negative except for a small
period at the beginning of the universe and thus, as the previous
case, a background filled with the interacting GGDE seems to be
unstable against perturbation. This implies that we cannot obtain
a stable GGDE dominated universe. However, the equations show a
sensitivity to the value of $\xi$ and we found that it is possible
to obtain a stable GGDE dominated universe but it may be in
contrast with theoretical values of $\beta$ \cite{caighost2}. One
important point is also the sensitivity of the instability to the
coupling parameter $b$. The larger $b$, leads to more instability
against perturbations. Thus once a gain we find a sign of
instability from this model in the presence of the interaction
term.
\section{Stability of the interacting GGDE in non-flat universe}\label{III}
\begin{figure}\epsfysize=5cm
{ \epsfbox{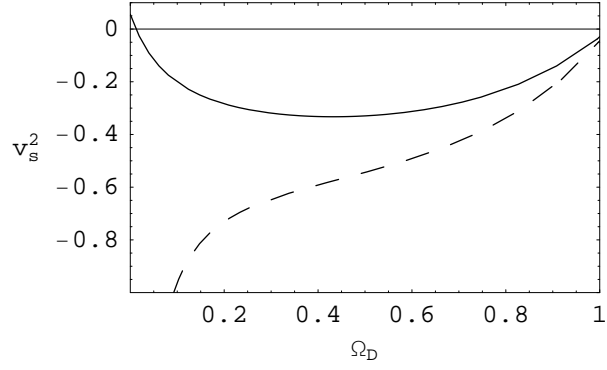}}\caption{The evolution of the
squared sound speed $v_s^2$ versus $\Omega_D$ for interacting GGDE
model in non-flat FRW universe. The solid curve corresponds to the
GGDE with $b=0.15$ and $\xi=0.1$, $\Omega_k=0.01$. Dashed curve
shows the original GDE model for $b=0.0.15$ .} \label{fvnfin}
\end{figure}

Although it is a general belief that the geometry of the universe
is flat, recently this assumption is challenged by several
observational evidences. For example, a closed universe is
marginally favored by observations from CMB \cite{nonflat2}.
Besides, the measurements of the cubic correction to the
luminosity-distance of supernova measurements support the idea of
a closed universe \cite{nonflat3}. Thus, there exist enough
motivations to consider DE problem in non-flat universe. Here, we
would like to extend the study of the GGDE  model to a universe
with special curvature. In such a case the first Friedmann
equations read
\begin{eqnarray}\label{Friedm}
H^2+\frac{k}{a^2}=\frac{1}{3M_p^2} \left( \rho_m+\rho_D \right),
\end{eqnarray}
where $k$ is the curvature parameter with $k = -1, 0, 1$
corresponding to open, flat, and closed universes, respectively.
Taking the energy density parameters (\ref{Omega}) into account
and defining the energy density parameter for the curvature term
as $\Omega_k=k/(a^2H^2)$, the Friedmann equation can be rewritten
in the form
\begin{equation}\label{fridomega2}
1+\Omega_k=\Omega_m+\Omega_D.
\end{equation}
Using the above equation the energy density ratio becomes
\begin{equation}\label{u2}
u=\frac{\rho_m}{\rho_D}=\frac{\Omega_m}{\Omega_D}=\frac{1+\Omega_k-\Omega_D}{\Omega_D}.
\end{equation}
The second Friedmann equation read
\begin{equation}\label{doth2}
\dot{H}=-4\pi G(P+\rho)+\frac{k}{a^2}.
\end{equation}
Following the previous sections one can easily see that the
equation of state parameter in the interacting non-flat GGDE may
be obtained as
\begin{equation}\label{wDn2}
w_D=-\frac{1}{2-\Omega_D-\xi}\left(2-\left(1+\frac{\xi}{\Omega_D}\right)\left(1+\frac{\Omega_k}{3}\right)+\frac{2b^2}{
\Omega_D} (1+\Omega_k)\right).
\end{equation}
From the second Friedmann equation, (\ref{doth2}), One can easily
obtain
\begin{equation}\label{q1n}
\frac{\dot{H}}{H^2}=-\Omega_k+\frac{3 }{2} \Omega_D[1+u+w_D],
\end{equation}
and therefore the deceleration parameter in this case is obtained
as
\begin{equation}\label{q1n}
q=-1-\frac{\dot{H}}{H^2}=-1-\Omega_k+\frac{3 }{2}
\Omega_D[1+u+w_D]
\end{equation}
Substituting Eqs. (\ref{u2}) and (\ref{wDn2}) in (\ref{q1n}) we
obtain
\begin{equation}\label{q2n}
q=\frac{1}{2}\left(1+\Omega_k\right)-\frac{3\Omega_D}{2(2-\Omega_D-\xi)}\left[2-\left(1+\frac{\xi}{\Omega_D}\right)\left(1+\frac{\Omega_k}{3}\right)+\frac{2b^2}{
\Omega_D} (1+\Omega_k)\right].
\end{equation}
Finally the evolution equation of the GGDE in a non-flat
interacting case can be written as
\begin{equation}\label{Omegaprime2n}
\frac{d\Omega_D}{d\ln
a}=3\Omega_D\left[\frac{\Omega_k}{3}+\frac{1-\Omega_D}{2-\Omega_D-\xi}\left(2-\left(1+\frac{\xi}{\Omega_D}\right)\left(1+\frac{\Omega_k}{3}\right)+\frac{2b^2}{
\Omega_D}
(1+\Omega_k)\right)-\frac{b^2}{\Omega_D}(1+\Omega_k)\right].
\end{equation}
In the limiting case $\Omega_k=0$, the results of this section,
restore their respective equations in flat FRW universe derived in
the previous sections and for $\xi=0$ the formulas of
\cite{shemov} are retrieved. The interested reader can see
\cite{ebrsheyggde} for detailed discussion about this case. Here
we consider main task of this section, study the stability of
interacting GGDE model in a universe with spacial curvature. From
the energy balance equation (\ref{consdi}) we can obtain
\begin{equation}\label{rrhodot2}
    \frac{\rho}{\dot{\rho}}=-\frac{1}{3H(1+w_D+\frac{b^2}{\Omega_D})}.
\end{equation}
Taking the time derivative of Eq. (\ref{wDn2}), yields
\begin{equation}\label{wdotnfint}
    \dot{w}_D=-\frac{\Omega_D^2(1-\frac{\Omega_k}{3})+\left(2b^2(1+\Omega_k)-\xi(1+\frac{\Omega_k}{3})\right)
    (2\Omega_D-2+\xi)}{{(2-\Omega_D+\xi)^2\Omega_D^2}}\dot{\Omega}_D
\end{equation}
Using the above relation, replacing (\ref{wDn2}) in Eq.
(\ref{rrhodot2}), and inserting them into (\ref{speeddef2}), after
using the time derivative version of (\ref{Omegaprime2n}) one gets
\begin{equation}\label{v2nfint}
    v_s^2=-2\frac{(\Omega_D-1)(\xi-\Omega_D)}{(\Omega_D-2+\xi)^2(\Omega_D+\xi)}
    +2b^2\frac{(3\Omega_D-4+\xi)(1+\Omega_k)}{(\Omega_D-2+\xi)^2(\Omega_D+\xi)}-
    \frac{2}{3}\Omega_k\frac{(\Omega_D+\xi)^2-\Omega_D-3\xi}{(\Omega_D-2+\xi)^2(\Omega_D+\xi)}.
\end{equation}
Setting $\Omega_k=0=b$ the above relation reduces to the flat
noninteracting respective relation. Also the squared sound speed
of the flat interacting case can be retrieved when $\Omega_k=0$.
In order to obtain an insight on the stability issue of the
interacting GGDE in a non-flat FRW universe, we should consider
the sign of $v_s^2$ during the evolution of the universe. To this
end we plot $v_s^2$ versus $\Omega_D$, where the value of
$\Omega_D$ indicates different epoches of the universe evolution.
Fig. \ref{fvnfin} shows the result which clearly indicates an
almost same behavior as the flat interacting case and it seems
that the presence of the curvature term does not lead to a
significant difference in the stability issue at least from this
feature (the squared sound speed). The squared sound speed,
($v_s^2$), is positive for $\Omega_D<\xi$ and is negative
otherwise, indicating the instability of the universe against
perturbations in the GGDE background. One can easily find from
(\ref{v2nfint}) that increasing $b$ will result more instability
in the universe. Finally we can say that the stability of a
universe filled with GGDE and matter against small perturbations
crucially depends on the value of $\xi$ parameter which is defined
on the base of the $\beta$ parameter in Eq. (\ref{GGDE}).
\section{Summary and Discussion} \label{sum}
Due to the lack of observational evidences every DE model which
can explain the current acceleration of the universe seems to be
acceptable. However, every new model accompanies by many
consequences which should be explored. In this paper we considered
an important feature, which every DE models should provide. Based
on cosmic observations our universe is in a stable DE dominated
epoch. In this paper we tried to see if the GGDE model is capable
to result in a stable dominated universe. The main point made the
authors to consider such an issue in the GGDE comes from the
result obtained in \cite{caighost2,ebrsheyggde}. In these papers
authors found that the second term in the GGDE energy density
(\ref{GGDE}), has a negative contribution to the energy density.
This negativity is in contrast of the first term and we thought
may such a behavior could result in some signs of stability in the
model. To this end, we used the squared sound speed
($v_s^2=\frac{dP}{d\rho}$) as the main factor for studying the
stability. If $v_s^2$ is positive the GGDE could be stable against
perturbations. When $v_s^2$ is negative we encounter signs of
instability in the background spacetime. We have discussed several
cases including whether there is or not an interaction between
dark matter and GGDE and whether there is or not a curvature term
in the background metric. We found that the stability of the GGDE
model crucially depends on the parameter $\xi$ and adjusting $\xi$
to suitable values the model is capable to result a stable DE
dominated universe. However, observational constraint on the GGDE
model \cite{caighost2} rejected mandatory range for $\xi$ to leave
a chance of stable GGDE dominated universe as the present state of
the universe. As a result, the universe filled with dark matter
and GGDE component cannot lead to a stable GGDE dominated
universe. We also observed that the instability of the interacting
GGDE increases with increasing the interacting coupling parameter
$b$. Thus, in comparison to the original GDE, the GGDE just has
theoretical chance to lead a GGDE dominated universe, however this
is ruled out by the present observational constraints. Readers
should note that a complete discussion on the stability issue
needs considering different features. Here, we just discussed the
sign of the square sound speed in different epoches which can be
taken as a sign of stability or instability of the model. Other
features of the stability will be addressed elsewhere.
\acknowledgments{This work has been supported financially by
Research Institute for Astronomy and Astrophysics of Maragha
(RIAAM) under research project  number No. 1/2782-63.}

\end{document}